\documentclass[a4paper,10pt]{article}
\usepackage{graphicx}

\begin{document}


\title{Implications of pc and kpc jet asymmetry to the cosmic ray acceleration\footnote{Published in International Journal of Modern Physics: Conference Series, Vol. 8 (2012), 331, World Scientific Publishing Company}}

\author{ Gizani$^1$, Nectaria A. B.\\
$^1$ Physics Laboratory, School of Science and Technology, Hellenic Open University, \\ Patra, Greece \\}

\date{}
\maketitle

\paragraph{Abstract}
We probe the role that the directional asymmetry, between relativistic outflows and kilo-parsec scale jets, play in the acceleration of cosmic rays. For this reason we use two powerful, nearby Active Galactic Nuclei (AGNs). These radio galaxies are atypical compared to the usual AGN as they contain ring-like features instead of hotspots. Our VLBI radio data have revealed a substantial misalignment between their small and large scale jets. Taking into account the overall information we have obtained about the AGNs themselves (VLA and VLBI radio data at 18 cm) and their clusters (X-ray observations) our study supports the present ideas of powerful radiogalaxies (radio quiet and radio loud) being sources of cosmic rays as well as their ability to accelarate the latter to ultra high energies.\\


\section{Introduction}

Ultra-high energy cosmic rays (UHECR) have been detected by the Pierre Auger, AGASA and HiRes experiments. UHECRs are cosmic rays with energies higher than the GZK cutoff\cite{gzka}--\cite{gzkb}, i.e. higher than a few $10^{19}$ eV. This cutoff comes (a) from proton primaries loosing energy while interacting with the Cosmic Microwave Background (CMB) by e$^{-}$, e$^{+}$ pair and by producing photopions (photomesons), and (b) as energetic nuclei are involved in photodisintegration. 

Using the Hillas criterion\cite{hillas} we can estimate the maximum energy $E_{\rm max} = Q\beta Bl$ that particles with charge $Q=Ze$ can acquire in an acceleration region of size $l$, magnetic field strength $B$, and velocity of magnetic field transport $v=\beta c$, taking into account Fermi first order acceleration. 

The HiRes data showed an isotropic distribution of the UHE particles (eg.~\cite{abbasi}). The Auger data (eg.~\cite{a1}--\cite{a2}) revealed arrival directions coming from the supergalactic plane (especially from the direction of the radiogalaxy Centaurus A), correlating the detected events with active galactic nuclei (AGN) within $\approx 100$ Mpc from the observer. The composition of the UHECRs is still unclear. HiRes suggests a proton composition (energy interval 10$^{18}-5\times 10^{19}$ eV), while fitting Auger data with hadronic model, heavier particles are favored (even iron nucleus) with energies 5$\times 10^{19}$ eV (eg. ~\cite{abreu}).

There is still a debate for the origin of the cosmic rays with energy values in the range 10$^{17}$ eV to 10$^{19}$ eV observed at Earth, as the Galaxy gives way to extragalactic sources (eg.~\cite{Berezinsky}): the Galactic magnetic field cannot confine cosmic rays at energies $\geq 10^{18.5}$ eV.
Potential candidates of the extragalactic UHECR sources and accelaration sites include gamma ray bursts (GRB) (eg.~\cite{murase}); supernovae and hypernovae (eg.~\cite{wang}); AGN (eg.~\cite{dremmer}--\cite{ensslin}), especially nearby powerful radio galaxies such as Centaurus A (eg.~\cite{ro}-- \cite{go}--\cite{honda}) and M87 (eg.~\cite{hc}-- \cite{mb}), including radio-quiet AGNs (eg.~\cite{asaf}); jets (and inner jets) (eg.~\cite{ptuskin}) in radio-loud (FRII AGNs\cite{FR}, and hence FRII-jets, eg.~\cite{dremmer}-- \cite{aharonian}) and radio-quiet AGNs (FRIs\cite{FR} and therefore FRI-jets, eg.~\cite{guo})~\footnote{ \cite{guv} suggests that the dichotomy between FRI and FRII jets could be due to environmental effects}; shocks (e.g. hotspots formed as the jets continuously feed the lobes with new material ~\cite{no}); radio lobes (eg.~\cite{hardcastle}); X-ray cavities (eg.~\cite{mathews},~\cite{guo},~\cite{guo2}); clusters of galaxies (eg.~\cite{sijacki}--~\cite{guo}) and magnetars (eg.~\cite{murase}). 

Jets are often considered as the main extragalactic sources of UHECRs (e.g.  \cite{lyutikov}). In that sense their kinetic energy should determine the maximum particle energy and shape the average source spectrum of the accelerated particles~\cite{ptuskin}. In addition relations (3) in ~\cite{lyutikov} show that HECRs are better accelerated at larger distances. AGN jets, spanning distances $\simeq$ 100 kpc are potential candidates. In particular as long as they stay relativistic while propagating to the intracluster medium, particles are accelerated without worrying about radiative loses.

In the current paper we adopt a cosmology in which $H_{\circ}$ = 65 km s$^{-1}$ Mpc$^{-1}$ and $q_{\circ}$ = 0.

\section{Observations and Results}

We have performed EVN phase referenced observations of Hercules A (18 mas resolution) and global VLBI observations of 3C\,310 (4 mas resolution) at 18~cm (see eg.~\cite{gizanie} and in preparation). Our motive was to resolve the parsec scale radio cores of these two powerful nearby, cluster hosts, FR1.5, ring-like featured radiogalaxies. Our observations revealed a substantially atypical pc-scale jet misalignment with respect to the kpc-scale ones. Her A is misaligned $\simeq 50^{\circ}$ to the line of sight (l.o.s.) implying $lorentz factor \sim$ 1.6 and $doppler factor \sim$ 1. The directional asymmetry is greater and more complex in 3C\,310. 

\section{Discussion}

We study Hercules A ($z=0.154$) and 3C\,310 ($z=0.054$) because they are quite atypical radiogalaxies in many senses. Their striking unusual similarity is that their radio lobes contain ring-like features instead of hotspots. Other closeby radio galaxies containing similar, but less distinct structures are for example DA240 ($z=0.035661$) and Centaurus A ($z= 0.001825$, Ekers, private communication). 

In this paper we discuss the ability of these AGNs and their jets to accelerate UHECRs. For this study we also use results by refs.eg.~\cite{gizanir},  \cite{gizanix},  \cite{leahya} and  \cite{nulsen} for Hercules A, and refs.eg.~\cite{van} and  \cite{miller} for 3C\,310 (radio and X-ray observations respectively). 

AGN jets, lobes, X-ray cavities and the Intracluster Medium (ICM) are all dynamically connected with eachother and independent studies so far have shown that they play an important role in the cosmic-ray production, acceleration and propagation. Jets are usually very collimated implying a process working very near the central engine. 

The eastern jet of Her A has the highest flux density of any jet found so
far. Its luminosity at 1.4 GHz is L$_{\rm east\,\rm jet} \sim 1.6 \times 10^{37}$ W. According to the Hillas criterion the maximum energy of accelerated particles in Her A would be $E_{\rm max} \propto L^{1/2}_{\rm jet}$. Less powerful is 3C\,310 compared to Her A.  Ref.eg.~\cite{lyutikov} adopting a cylindrically collimated jet model of powerful AGNs (i.e., FR II radiogalaxies, radio loud quasars and high power BL Lacs) have found that UHECR in the form of mostly protons can be accelerated in low spacial density sources from sub-parsec to hundreds of kiloparsec scales as radiative losses are negligible there, as long as the motion of the jet stays relativistic.  However radio quiet AGNs and their weak parsec-scale jets can also be sources of UHECRs as long as the cosmic rays consist of heavy nuclei \cite{asaf}. In the context of parsec-scale potential for acceleration ref.eg.~\cite{ensslin96} have proposed that subparsec scale acceleration is efficient as long it is comparable to the scale of jet generation or initial collimation. 

Our new VLBI observations have shown a substantial misalignment between parsec and kiloparsec jets of both sources, which could be supporting the unification suggestion that radiogalaxies are the unbeamed counterparts of quasars (eg.~\cite{Saripalli}). Moderate mas (milliarcsecond) asymmetry to the kiloparsec scale jet and similar to Her A misalignment to the l.o.s. has also been found in another nearby radiogalaxy~\cite{horiuchi}. This is Centaurus A,  the Auger experiment's favorite source of UHECRs. The magnetic field in the lobes of Cen A is of the order of a few $\mu$G like Her A (see ref eg.~\cite{gizanir} and in preparation). Although the received flux from nearby radiogalaxies with their closer jet pointing away from earth is small, they can also be sources of UHECRs. This can be done through deflection of CRs in the radio lobes of radiogalaxies like Cen A~\cite{dremmer}. Ref.eg.~\cite{lyutikov} point out that as long as the particle remains inside a jet (carrying large scale magnetic fields), the acceleration becomes theoreticaly maximum at an inverse gyro-frequency, implying that the most efficient acceleration occurs right before the particle leaves the jet.  

Steepening of radio emission suggests cooling of particles. Steep spectra in both sources imply short lifetimes and reacceleration of particles. The short cooling time of the emitting cosmic ray electrons (visible also in Her A~\cite{gizanir} and 3C310~\cite{van}) and the large extent of the radio sources suggest an ongoing acceleration mechanism in the ICM. 

ICM is prevented from a cooling catastrophically through various heating sources such as cavities, thermal filaments, and weak shocks detected in the X-rays. The radial elongation found in cavities indicates that the source of cosmic rays is a jet emanating from the central AGN~\cite{guo}, although the relation between cavity formation and AGN lobes has also been studied~\cite{guo}. Combined observations of thermal filaments and radio lobes can be used to trail the propagation of cosmic rays.  Fossil AGN jets and cocoons can act as reservoirs of CRs even if the AGN is 'dead'\cite{bp}. 

Ref.eg.~\cite{nulsen} have made deep Chandra observations of the Her A cluster. They have detected cavities which are not in the direction of the radio lobes. The authors have also found an enhanced ridge of nonthermal emission in a relatively cool and dense gas, extending $\sim 30^{\circ}$ from the north west to south east direction and at right angles to the cavities' axis. According to our VLBI observations the ridge is towards the direction of the parsec scale jets suggesting displacement of the gas by the inner jets while they expand to the ICM becoming mildly relativistic.

\end{document}